\documentclass[twocolumn,showpacs,preprintnumbers,amsmath,amssymb]{revtex4}

\usepackage{graphicx}
\usepackage{dcolumn}
\usepackage{bm}
\usepackage{dsfont}
\usepackage{mathrsfs}
\usepackage{amsmath}				

\newcommand{\be}{\begin{equation}}
\newcommand{\ee}{\end{equation}}
\newcommand{\bey}{\begin{eqnarray}}
\newcommand{\eey}{\end{eqnarray}}

\begin{document} 
\title{Properties of canonical determinants and a test of fugacity expansion \\ 
for finite density lattice QCD with Wilson fermions}
\author{Julia Danzer$^{a,\,b}$} 
\author{Christof Gattringer$^{a,\,c}$}
\affiliation{
\vspace{3mm}
$^a$Institut f\"ur Physik, FB Theoretische Physik,
Universit\"at Graz, 8010 Graz, Austria \\
$^b$Wegener Center for Climate and Global Change,
Universit\"at Graz, 8010 Graz, Austria  \\
$^c$Institute for Nuclear Theory,
University of Washington, Seattle, WA 98195, USA 
\vspace{2mm} 
}

\date{April 5, 2012}

\begin{abstract}
\vspace{4mm}
We analyze canonical determinants, i.e., grand canonical determinants projected to
a fixed net quark number. The canonical determinants are the coefficients in a
fugacity expansion of the grand canonical determinant and we evaluate them as the
Fourier moments of the grand canonical determinant with respect to imaginary chemical
potential, using a dimensional reduction technique. The analysis is done for two
mass-degenerate flavors of Wilson fermions at several temperatures below and above
the confinement/deconfinement crossover. We discuss various properties of the
canonical determinants and analyse the convergence of the fugacity series for
different temperatures.
\end{abstract}

\preprint{INT-PUB-12-016}

\maketitle

\section{Introduction}

QCD with finite density is an important topic of both, experimental and theoretical
studies. In particular one would like to understand the various  transitions to
different states of matter that are conjectured for the QCD phase diagram. Exploring these
phase transitions is clearly a non-perturbative problem and non-perturbative
methods must be applied for such studies. In principle the lattice QCD formulation
provides a suitable non-perturbative framework, and as long as the chemical
potential is zero this approach may be used to obtain reliable quantitative
results. 

However, if one works at finite density a severe technical problem emerges: The
fermion determinant is complex at non-zero chemical potential $\mu$
and cannot be used as
a probability weight in a Monte Carlo calculation. Various reweighting and
expansion techniques have been explored in recent years and the reviews at the
annual Lattice conference provide a glimpse at the corresponding developments
\cite{reviews}.   An interesting option is to work not in the grand canonical
approach with a chemical potential $\mu$, but instead to use canonical
partition sums with fixed net  quark number $q$. Several studies of this
alternative perspective can be found in the literature \cite{hasenfratz} --
\cite{BiDaGaLaLi}. So far the focus was mainly either on reweighting techniques or
full canonical simulations at a fixed quark number. 

Here we explore the perspectives of a fugacity expansion of the grand canonical
ensemble. The expansion coefficients are fermion determinants projected to a
fixed net quark number, so-called {\it canonical determinants}. The expansion is a
Laurent series in the fugacity parameter $e^{\mu \beta}$ ($\beta$ is the inverse
temperature) which has properties different from the more conventional Taylor
expansion in $\mu \beta$. However, the possible  application to lattice QCD at finite
density is the same for both expansions: One computes the expansion coefficients in
simulations at $\mu = 0$  and uses the series to explore QCD at $\mu > 0$. 

In this article we discuss the evaluation and the properties of canonical
determinants. They are computed as Fourier moments of the fermion determinant at
imaginary chemical potential $\mu \beta = i \varphi$, 
which we evaluate  efficiently using a recently proposed
dimensional reduction formula \cite{DaGa}. We study the distribution  properties of the
canonical determinants and analyze the convergence of the fugacity expansion for various
temperatures in the confined and deconfined phases. This analysis is done for a fixed numerical 
effort, i.e., a fixed number of 256 sampling points of $\varphi$ in the interval $(-\pi,\pi]$ used for the evaluation of the Fourier transformation to obtain the canonical determinants. 

\section{Canonical determinants and their evaluation}

The starting point for our study of canonical determinants and the fugacity
expansion is Wilson's lattice Dirac operator $D(\mu)$ with chemical  potential
$\mu$ on lattices of size $N_s^3 \times N_t$. The lattice spacing will be denoted
by $a$, such that $\beta = a N_t$. For two mass-degenerate flavors the
corresponding  grand canonical partition function is given by
\begin{equation}
Z(\mu) \; = \; \int {\cal D}[U] \, e^{-S_G[U]} \, \det[D(\mu)]^2 \; ,
\label{zgrand}
\end{equation}
where $S_G[U]$ is the action for the gauge fields $U$, ${\cal D}[U]$ the
path integral measure and $\det[D(\mu)]$ the fermion determinant
for chemical potential $\mu$, which in the following will be referred to as 
{\it grand canonical determinant}. 
The grand canonical determinant can be expanded in a 
finite fugacity series, 
\begin{equation}
\det[D(\mu)] \; = \; \sum_q e^{\mu\beta q} \, D^{(q)} \; ,
\label{detfugacity}
\end{equation}
where the sum runs over integer valued quark numbers $q \in [-6N_s^3,+6N_s^3]$. 
The expansion coefficients $D^{(q)}$ are the 
canonical determinants and may be obtained using Fourier transformation
with respect to an imaginary chemical potential,
\begin{equation}
D^{(q)} \; = \; \frac{1}{2\pi} \int_{-\pi}^\pi \!\!\!\! d \varphi  \;
e^{-i q \varphi}\, \det[D(\mu\beta = i\varphi)] \;\; .
\label{DQfourierdef} 
\end{equation} 
While for real chemical potential $\mu$ the grand canonical determinant is complex,
for imaginary chemical potential the determinant $\det[D(\mu\beta = i\varphi)]$
is real. One may explore the generalized 
$\gamma_5$-hermiticity relation for Wilson fermions, 
$\gamma_5 D(\mu) \gamma_5 = D(-\mu)^\dagger$,  
to establish the relation $D^{(-q)} \, = \, ( D^{(q)} )^*$
between canonical determinants with positive and negative quark numbers
$q$. For vanishing quark number  $q = 0$ the canonical determinant is real, 
i.e., $D^{(0)} \in \mathds{R}$.

\begin{table}[b]
\begin{center}
\begin{tabular}{ccccc}
\quad  $\beta_{gauge}\; $ \quad & \qquad $\kappa$ \qquad \quad & \quad $a$ [fm]$\;\;$ \quad & 
\quad  $T$ [MeV]$\;\;$  \quad & \quad  $q_{cut}\;\;$ \quad  \\
\hline
5.00 & 0.158 & 0.343(2) & 144(1) & 12 \\
5.10 & 0.158 & 0.322(4) & 153(2) & 13 \\
5.15 & 0.158 & 0.313(3) & 157(2) & 13 \\
5.20 & 0.158 & 0.300(1) & 164(1) & 14 \\
5.25 & 0.158 & 0.284(5) & 173(3) & 16 \\
5.30 & 0.158 & 0.260(1) & 189(1) & 21 \\
5.35 & 0.158 & 0.233(2) & 211(2) & 25
\end{tabular}
\end{center}
\vspace{-3mm}
\caption{Parameters of our numerical simulation.}
\label{paramstable}
\end{table}

In principle Eq.~(\ref{DQfourierdef}) is a very elegant expression for the
canonical determinants $D^{(q)}$. However, the Fourier  integral has to be
evaluated numerically and the evaluation of the grand canonical determinant
$\det[D(\mu\beta = i\varphi)]$ for the necessary  $\varphi$-values is
computationally rather costly. To alleviate the problem we  use a domain
decomposition technique \cite{DaGa} where the grand canonical fermion determinant
may be rewritten exactly in a dimensionally reduced form,
\begin{equation}
\det[D(\mu)]  =  A_0 \det 
\big[ \mathds{1} - H_0 - e^{\mu \beta} H - e^{-\mu \beta} H^{\, \dagger} \big] \;
,
\label{detH0Hpm}
\end{equation}
which holds both for real and imaginary chemical potential.  Here $A_0$ is a real
factor which depends only on the background gauge field configuration but is
independent of the chemical potential $\mu$. $H_0 = H_0^\dagger$, and  $H$ are
matrices that are built from propagators on  sub-domains of the lattice and live on
only a single time slice  (see \cite{DaGa} for details). Thus the determinant in
(\ref{detH0Hpm}) is dimensionally reduced, i.e., the determinant is taken over a 
matrix with $12 N_s^3$  rows and columns, where the factor 12  comes from the color
and Dirac indices. The terms $H_0$ and $H$ can be completely  stored in memory and
are then used  many times for the evaluation of $\det[D(\mu\beta = i\varphi)]$ at
the necessary values of $\varphi$. Due to the reduced size of the matrix where the
determinant has to be evaluated, we gain a factor of ${\cal O} (N_t^3)$, which
here, at $N_t = 4$, corresponds to a speedup by a factor of 64. The Fourier
integral (\ref{DQfourierdef}) is evaluated numerically with standard techniques
\cite{numrec} using  256 values of $\varphi$ in the interval $(-\pi,\pi]$.  We
remark at this point that also alternative dimensional reduction formulas with
different properties were proposed \cite{hasenfratz,altfactstaggered,adams,altfactwilson}.
 
Our numerical tests are done on dynamical configurations with two flavors of Wilson
fermions at $\mu = 0$ generated with the MILC code \cite{milc}.  In this
exploratory study of the canonical determinants and the fugacity  expansion we are
here limited to a single lattice volume, $8^3 \times 4$. The parameters (inverse
gauge coupling $\beta_{gauge}$ and hopping parameter $\kappa$) and the values for the corresponding
lattice spacing $a$ and the temperature $T$ are taken from  \cite{wilsonparams} and
listed in Table \ref{paramstable}. Due to the small lattice the pion mass has to be
kept rather large -- it is close to 950 MeV for all our ensembles. For each set of
parameters  we use a statistics of 200 configurations. All errors we show are
statistical errors determined with the jackknife method. In addition to the
parameters of the simulation, in  Table~\ref{paramstable} we also list the cutoff
parameter $q_{cut}$ used in  the fugacity expansion which we discuss in more
detail in Section IV.   
 
In the range of temperatures we consider, the lowest three values of $T$ are in the
confined phase, the crossover into the deconfined phase is at $T \sim 165$
MeV for our simulation parameters, and the two largest temperatures are in the deconfined phase.

\begin{figure}[t]
\begin{center}
\hspace*{-1mm}
\includegraphics[width=80mm,clip]{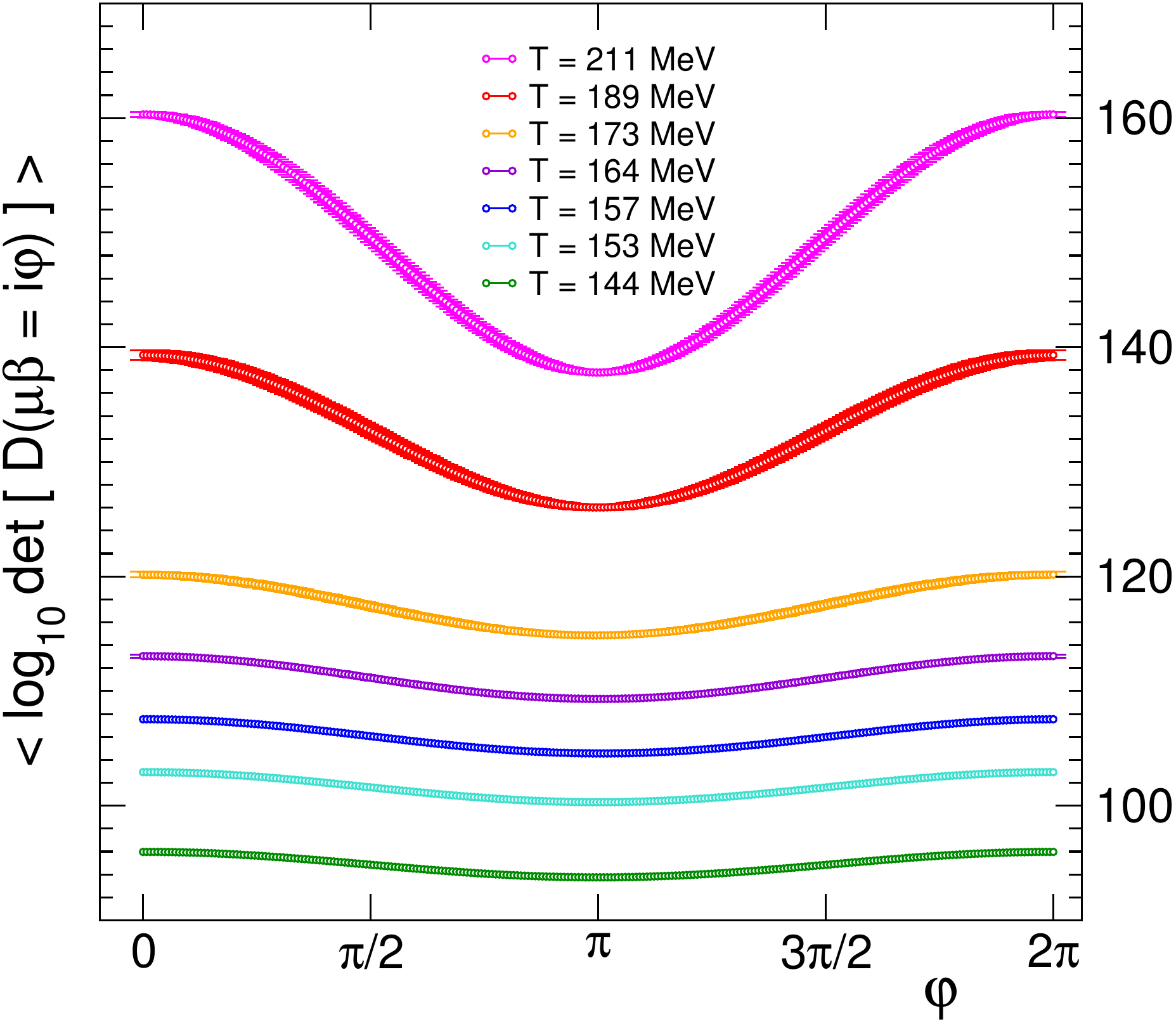} 
\end{center}
\vspace{-3mm}
\caption{Average of the logarithm of the grand canonical 
fermion determinant as a function of the boundary phase 
$\varphi$. We compare the results for different temperatures from 144 MeV (bottom curve) to 211 MeV (top). \hfill} 
\label{fulldet_vs_phi} 
\end{figure}

As a first result we now have a look at the average  $\langle \log_{10} \det[D(\mu\beta
= i\varphi)] \rangle$ of the logarithm of the grand canonical determinant as a 
function of the imaginary chemical potential parameter $\varphi$. This is interesting 
because the Fourier moments with respect to $\varphi$ are the canonical determinants
 we want to study and the $\varphi$-dependence of $ \det[D(\mu\beta = i\varphi)]$ 
 already provides a first insight into their properties.  $\langle \, .. \,
\rangle$  is the average over gauge configurations generated for two mass 
degenerate  flavors of Wilson fermions at $\mu = 0$ as detailed above. We study the
$\varphi$-dependence of  $\langle \log_{10}
\det[D(\mu\beta = i\varphi)] \rangle$ using all 256 values of $\varphi$ in the
interval $(-\pi,\pi]$. The integrands $\det[D(\mu\beta = i\varphi)]$
are computed with the dimensional  reduction formula (\ref{detH0Hpm}) as outlined. 
We show the corresponding results for all our temperature values  in
Fig.~\ref{fulldet_vs_phi}.  

The figure clearly demonstrates that in the confined phase the grand canonical 
determinant shows only a weak dependence on $\varphi$, while above the crossover  a pronounced variation with $\varphi$ develops. This behavior has a simple and well
known interpretation: An imaginary chemical potential $\mu \beta = i \varphi$ is
equivalent to an additional temporal boundary condition for the fermion fields,
$\psi(\vec{x},N_t+1) = -  e^{i\varphi} \psi(\vec{x},1)$, where the minus sign is
the usual anti-periodic temporal  boundary condition for fermions. While in the confined phase
the correlations are  short ranged and do not give rise to a dependence on the
boundary angle $\varphi$,  in the deconfined phase correlations over distances
larger than the temporal extent of the lattice generate the non-trivial response to
changing $\varphi$. The canonical determinants $D^{(q)}$ are the Fourier moments of
the $\varphi$-dependence and the first analysis in Fig.~\ref{fulldet_vs_phi}
already demonstrates that in the deconfined phase the higher modes, i.e., $D^{(q)}$
with larger values of $|q|$ will play a more prominent role.

\begin{figure*}[t]
\begin{center}
\includegraphics[width=160mm,clip]{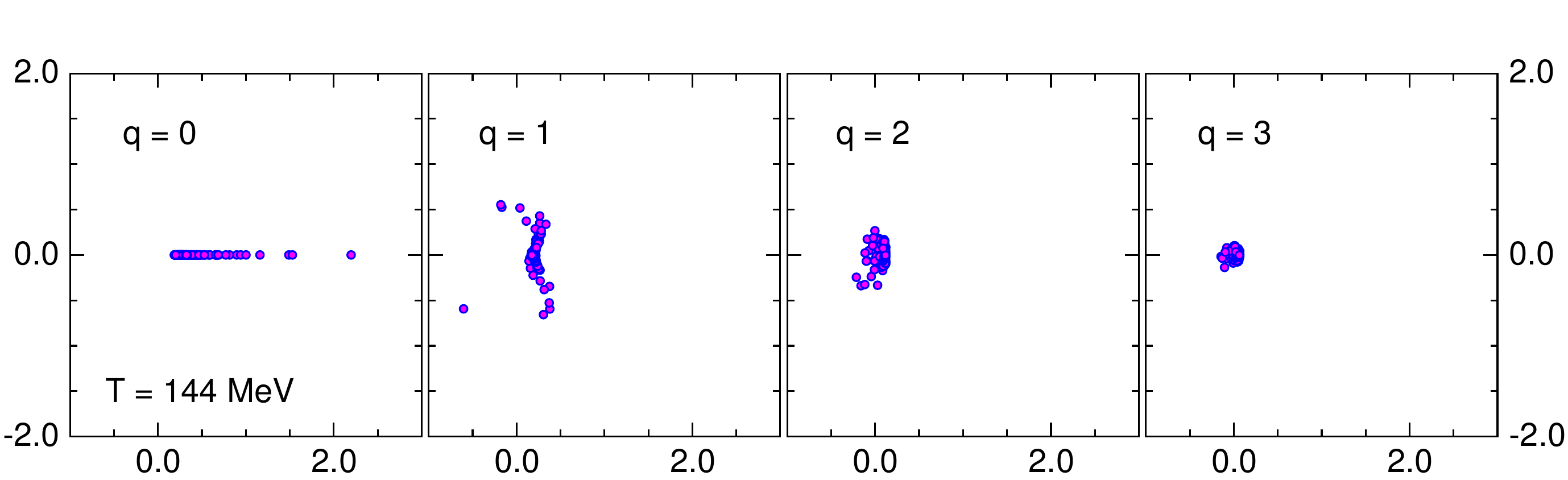} 
\vskip3mm
\includegraphics[width=160mm,clip]{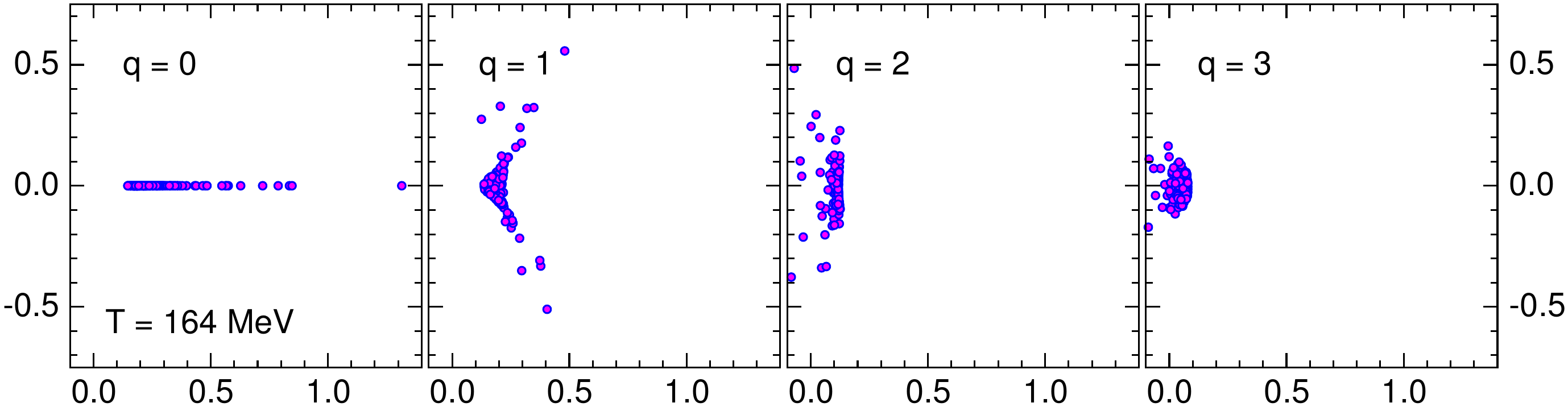} 
\vskip3mm
\includegraphics[width=163.5mm,clip]{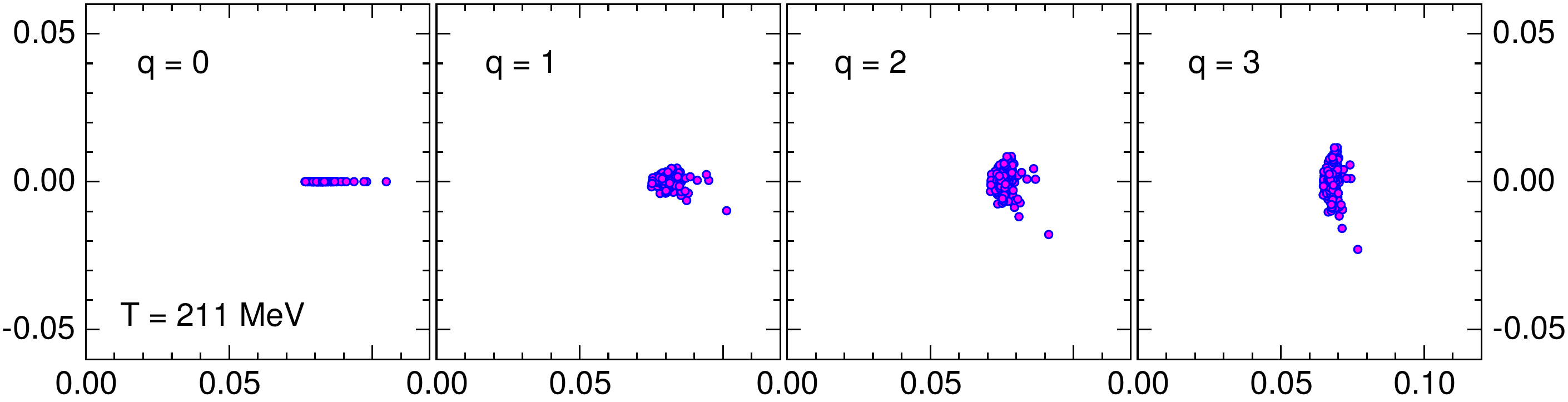} 
\end{center}
\vspace{-3mm}
\caption{Scatter plot of the canonical determinants $D^{(q)} / \det[D(\mu = 0)]$  in the
complex plane at $T = 144$ MeV (top row of plots), $T = 164$ MeV (middle) and $T = 211$ MeV (bottom) for $q = 0,1,2,3$. The canonical determinants are
normalized by the grand canonical determinant at $\mu = 0$. Note that we use different
scales for the different temperatures. \hfill} 
\label{scatter}
\end{figure*}

\section{Properties of the canonical determinants} 

Having discussed the canonical determinants $D^{(q)}$ and the setting of our
calculation we now come to analyzing properties of the $D^{(q)}$. In
Fig.~\ref{scatter} we show scatter plots in the complex
plane for  $D^{(q)} / \det[D(\mu = 0)]$, i.e.,  the canonical determinants
normalized by the grand canonical determinant at $\mu = 0$. This normalization is such that when one sums over $q$ all data points $D^{(q)} / \det[D(\mu = 0)]$ in the complex plane for a single configuration the result is $1$.  In the plot we show results for $q = 0,1,2,3$ and compare $T = 144$ MeV (top row of plots in Fig.~\ref{scatter}), $T = 164$ MeV (middle) and $T = 211$ MeV (bottom). Note that different axis scales are used. 
 
As already discussed, for $q=0$ the canonical determinants $D^{(q=0)}$ must be
real, a fact that is obvious in the figures. The corresponding values scatter on
the real axis  for all temperatures.  For $q > 0$ the points spread out also in the imaginary direction. In the confined phase ($T = 144$ MeV) the points for the higher Fourier modes $D^{(q)}$ then
quickly  approach the origin of the complex plane when increasing $q$. The
situation is different in the deconfined phase  ($T = 211$ MeV), where we observe
that for increasing $q$ the values of $D^{(q)}$ do not move toward the origin as quickly as for $T = 144$ MeV. For $T = 164$ MeV (middle row of plots in Fig.~\ref{scatter}) one finds an intermediate behavior. The relative size of the 
canonical determinants $D^{(q)}$ at the different temperatures reflects the observation we have already made in the discussion of
Fig.~\ref{fulldet_vs_phi}: For temperatures above the crossover transition higher
Fourier components $D^{(q)}$ are necessary to resolve the strong dependence of the 
grand canonical determinant $\det[D(\mu\beta = i\varphi)]$ on the boundary phase
$\varphi$. 

We conclude the discussion of Fig.~\ref{scatter} with noting, that the  scatter plots of
$D^{(q)}$ for the smaller values of $q$ show a curious pattern, in particular the $q = 1$ data for $T = 144$ and 164 MeV: They arrange in two oblong structures that seem to have a preferred angle relative to the real axis. A similar pattern at temperatures near the crossover was observed also in canonical determinants for staggered fermions \cite{unpublished}. At the moment we do not understand the origin of this pattern, but it might be related to the properties of the $D^{(q)}$ under center transformations which rotate them according to  $D^{(q)} \, \rightarrow \, z^{q\!\!\mod\!3} D^{(q)}$, where $z \in \{1, e^{i2\pi/3}, e^{-i2\pi/3}\}$ is the center element multiplied to all temporal gauge links in a fixed time slice. For pure gauge theory the spontaneous breaking of center symmetry is intimately related to the deconfinement transition, and it could be that the pattern we observe is a remnant of this mechanism. 
This speculation is further supported by the observation, that the phase of the Polyakov loop (the order parameter for confinement in pure gauge theory) is strongly correlated with the phases of the $D^{(q)}$ \cite{unpublished}.
Also in an analysis of the canonical determinants of the quenched case \cite{DaGa2,BiDaGaLaLi} it was demonstrated that the $D^{(q)}$ in the broken phase very cleanly map the center orientation of the underlying center sector.

\begin{figure}[t]
\begin{center}
\hspace*{-10mm}
\includegraphics[width=70mm,clip]{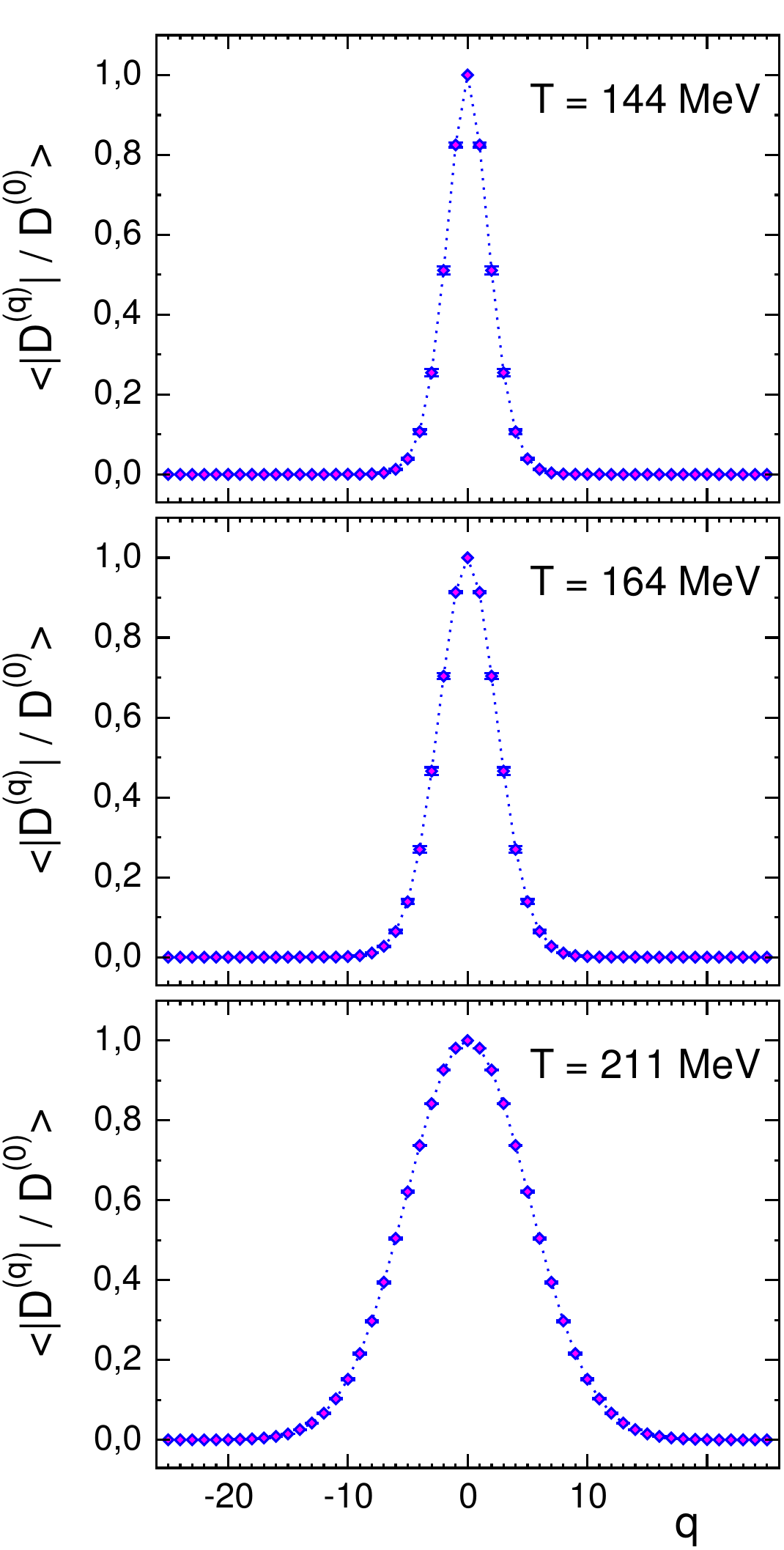} 
\end{center}
\vspace{-5mm}
\caption{Distribution of $\langle | D^{(q)} | / 
D^{(0)} \rangle$ as a function of the quark number $q$. We compare three temperatures: $T =
144$ MeV (top plot), $T = 164$ MeV (middle) and $T = 211$ MeV (bottom). 
\hfill} 
\label{Dq_dist} 
\end{figure}

\begin{figure}[t]
\begin{center}
\vspace*{2mm}
\hspace*{-1mm}
\includegraphics[width=80mm,clip]{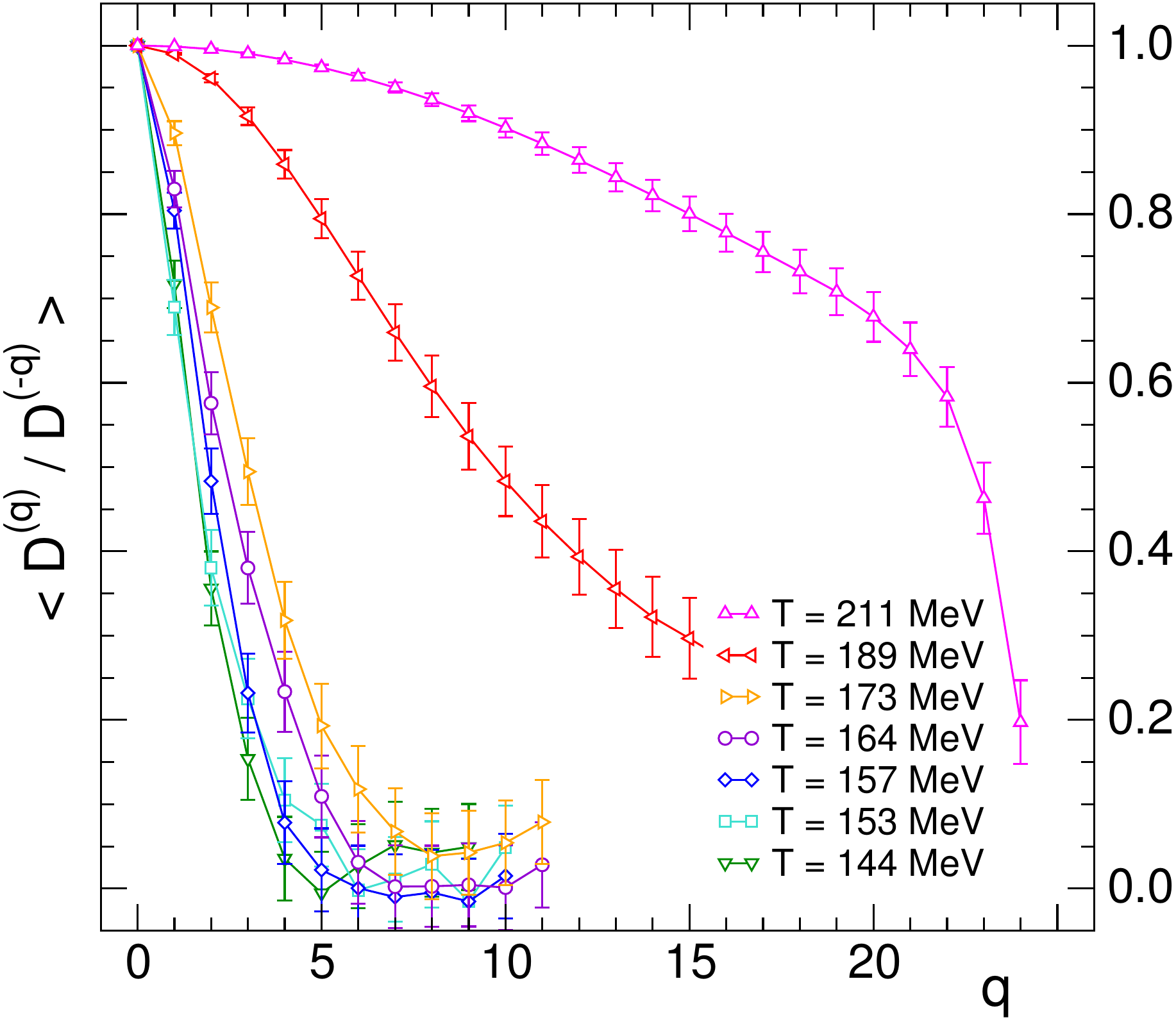} 
\end{center}
\vspace{-3mm}
\caption{Expectation value of the phase of the canonical determinants,
$\langle e^{i2\theta^{(q)}}\rangle = \langle D^{(q)}/D^{(-q)}\rangle$ as a function of $q$. We compare the results for different temperatures. \hfill} 
\label{Dq_phaseaver} 
\end{figure}

The change of the relative weight of the $D^{(q)}$ with temperature is manifest
also in Fig.~\ref{Dq_dist}, where we study the size distribution  $\langle |
D^{(q)} | / D^{(0)} \rangle$ of the canonical determinants normalized to the
trivial $q = 0$ sector as a function of $q$ and again compare results 
for $T = 144$ MeV (top), $T = 164$ MeV (middle) and $T = 211$ MeV (bottom). 
The plots show that the distribution roughly
follows a Gaussian
centered around the dominant   $q = 0$ sector. While in the confined phase the
distribution  is rather narrow, in the deconfined phase it widens and reflects the
fact that at higher temperatures also sectors with larger quark numbers become
accessible.  When one turns on a chemical potential $\mu > 0$ the canonical determinants are multiplied with powers of the fugacity factor, i.e., $D^{(q)} \rightarrow e^{\mu \beta q} D^{(q)}$, which has the effect of shifting the center of the distribution towards larger quark numbers.

Finally, in Fig.~\ref{Dq_phaseaver} we have a look at the phase $\theta^{(q)}$ of the canonical determinants $D^{(q)}$. More precisely we look at $\langle e^{i 2 \theta^{(q)}} \rangle =  \langle D^{(q)}/ D^{(-q)} \rangle$ (remember that $D^{(-q)} = {D^{(q)}}^\star$). In Fig.~\ref{Dq_phaseaver} we show   $\langle D^{(q)}/ D^{(-q)} \rangle$ as a function of $q$ for different values of the temperature. 

Below the crossover temperature the results for $\langle D^{(q)}/ D^{(-q)} \rangle$ drop to 0 rather quickly with 
increasing $q$, i.e., the phases of the $D^{(q)}$ for $q$ above roughly $q = 5$ fluctuate strongly and average to a 
very small number. For the largest temperatures the decrease of   $\langle D^{(q)}/ D^{(-q)} \rangle$ with $q$ is
slower and the phases have a sizable expectation value up to $q \sim 25$ (we stress that these are statements
specific to the parameters of our calculation, in particular the volume used). 

The decrease of the $\langle D^{(q)}/ D^{(-q)} \rangle$ with increasing $q$ also sheds light on how the fermion sign
problem may be viewed in the fugacity expansion: When increasing $\mu$ the powers $e^{\mu \beta q}$ of the fugacity 
factors  put a larger weight on the canonical determinants $D^{(q)}$ with higher quark numbers, which in turn
are the ones that are suppressed due to the fluctuations of their phase. The fact that for higher temperatures the
sign problem is milder is clearly visible from the slower decrease of $\langle D^{(q)}/ D^{(-q)} \rangle$ for larger
temperatures.

\begin{figure*}[t]
\begin{center}
\includegraphics[width=170mm,clip]{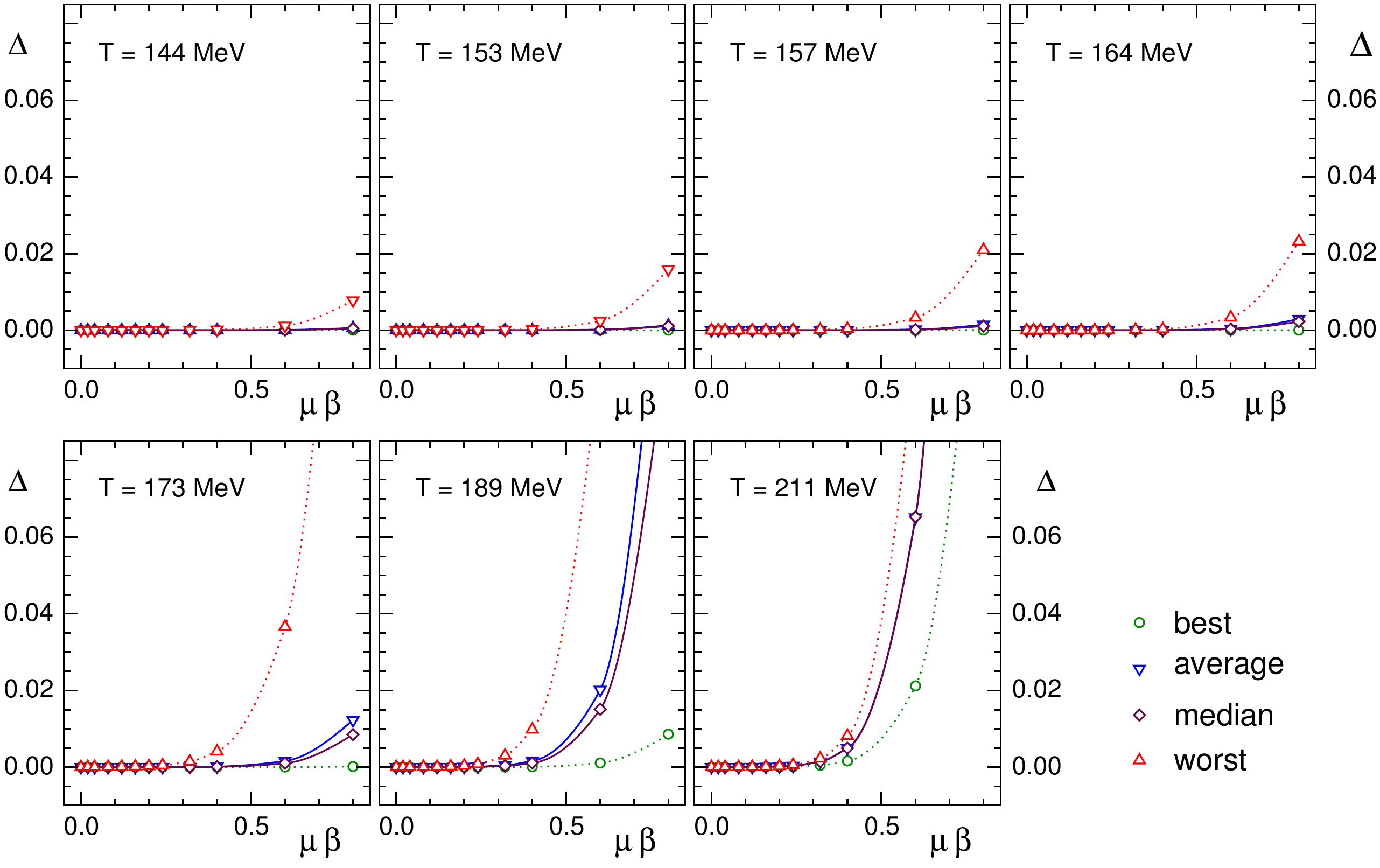} 
\end{center}
\vspace{-3mm}
\caption{Relative error $\Delta$ (see the text for its definition) of the 
fugacity expansion as a function of $\mu \beta$ at different temperatures. We show
the error of the best configuration, the worst configuration, the average over all
configuration as well as the median. \hfill} 
\label{fugacitytest}
\end{figure*}

\relax

\section{Convergence properties of the fugacity series}

After the study of the canonical determinants $D^{(q)}$, we now can start to analyze the
fugacity series (\ref{detfugacity}). In a numerical study it is necessary to truncate the
series and we denote the truncated fugacity expansion by
\begin{equation}
S(\mu)_{q_{cut}} \; = \; \sum_{q = -q_{cut}}^{q_{cut}} \, e^{\mu \beta q} \, D^{(q)} \; ,
\label{truncexpand}
\end{equation}
where $q_{cut}$ denotes the highest (anti-) quark sector we take into account in the
truncated series. 

The central question we want to study in this section is how well the truncated
series $S(\mu)_{q_{cut}}$ approximates the full grand canonical determinant
$\det[D(\mu)]$ and how the quality of the approximation depends on the parameters
$\mu$ and $T$.  This is a question which we address both for individual
configurations and for the gauge average. We stress at this point, that for a
finite volume and non-zero temperature, the representation of $\det[D(\mu)]$ by
$S(\mu)_{q_{cut}}$ would be exact with  $q_{cut} = 6N_s^3$  if all coefficients
$D^{(q)}$ were known with arbitrary precision. Thus as long as the lattice is
finite such that no singularities can emerge, an insufficient representation of
the grand canonical determinant is only due to limited accuracy in the numerical
evaluation of the Fourier integrals (\ref{DQfourierdef}).

We assess the quality of the approximation by considering the relative error
\begin{equation}
\Delta \; = \; \left| \frac{ S(\mu)_{q_{cut}}  - \det[D(\mu)] }{ S(\mu)_{q_{cut}} 
\, } \right| \; \; .
\label{Deltadef}
\end{equation}
For the study of $\Delta$ we also evaluated
the grand canonical determinant $\det[D(\mu)]$ for a few values of real chemical potential,
again applying the dimensional reduction formula (\ref{detH0Hpm}).
These values are then used for the assessment of the fugacity expansion using $\Delta$. 

Inspecting the fugacity series (\ref{detfugacity}) or (\ref{truncexpand}) one
observes that for increasing $q$ there is competition of two terms: The powers of
the fugacity, i.e., the factors $e^{\mu \beta q}$, are terms that grow
exponentially with $q$ and their rate of growth is determined by the product $\mu
\beta$. The exponential growth with $q$ is compensated by the quick decrease of the
canonical determinants $D^{(q)}$. This decrease is roughly Gaussian as we 
demonstrated in Fig.~\ref{Dq_dist} and thus dominates the increasing function
$e^{\mu \beta q}$. However, this mechanism can be spoiled by numerical
instabilities: If the higher coefficients $D^{(q)}$ are not known with perfect
accuracy (as will always be the case when they are evaluated via numerical Fourier
transformation), their fluctuations will be amplified by the factors $e^{\mu \beta
q}$. This effect is manifest in the behavior of $S(\mu)_{q_{cut}}$ when $q_{cut}$
is varied:  First $S(\mu)_{q_{cut}}$ quickly saturates as a function of $q_{cut}$,
but for too large values, e.g., already $q_{cut} \sim 30$ for some of our parameter
values, the series $S(\mu)_{q_{cut}}$ starts to diverge. Fortunately the $D^{(q)}$
decrease very rapidly: Fig.~\ref{Dq_dist} shows that also for our highest
temperature ensemble all $D^{(q)}$ with $|q| > 20$ essentially vanish, and the
series $S(\mu)_{q_{cut}}$ can be truncated at small $q_{cut}$. The optimal values
for $q_{cut}$ were determined using the relative error $\Delta$ and we list them in
Table~\ref{paramstable}.     
 
In Fig.~\ref{fugacitytest} we show how the relative error $\Delta$ of the fugacity
expansion depends on the parameters $\mu \beta$ and $T$. For all
200 configurations in each of our ensembles we compute the canonical determinants
$D^{(q)}$, as well as the canonical determinant $\det[D(\mu)]$ for several real values
of $\mu$. For these values of $\mu$ we can evaluate $\Delta$. 

Fig.~\ref{fugacitytest} shows $\Delta$ as a function of the dimensionless
combination $\mu \beta$, and we compare the results for all available temperatures.
For each ensemble   we show $\Delta$ for the configuration where $\Delta$ is
largest ("worst"), for the configuration where $\Delta$ is smallest ("best"), for
the median of the 200 configurations, as well as the average of $\Delta$ over all
configurations.     

From Fig.~\ref{fugacitytest} it is obvious that for the lower values of $T$ the
fugacity expansion has better convergence properties: At $T = 144$ MeV all four
categories (best,  worst, median and average) have a relative error $\Delta$
smaller than 1 \% for values of the chemical potential 
up to $\mu \beta = 0.8$ and the average error over all configurations remains 
below 1\% even up to $\mu \beta = 1.1$. For $T = 211$ MeV an
error smaller than 1 \% can be maintained only up to $\mu \beta = 0.4$. 
The reason is that here the distribution of the $D^{(q)}$ is much wider 
(compare Fig.~\ref{Dq_dist}), and  the $D^{(q)}$ for larger values of 
$|q|$ contribute significantly 
in the fugacity sum already at not too large $\mu \beta$. Since these higher 
Fourier modes are only known with less relative accuracy, the fugacity expansion
at the highest temperatures breaks down already at smaller chemical potentials. 

We stress that our convergence analysis is specific for the setting
we use here, i.e., lattice size $8^3 \times 4$, parameters as listed in
Table~\ref{paramstable} and  256 values of $\varphi$ in the evaluation of the
Fourier integrals (\ref{DQfourierdef}). In particular increasing
the number of sampling points for $\varphi$ in the Fourier integral will allow for a higher
precision of the $D^{(q)}$ which leads to better convergence properties of the
fugacity sum, such that higher values of $\mu \beta$ can be reached. 

An interesting question is how the number of $D^{(q)}$ that need to be evaluated
depends on the volume of the box. This number is roughly proportional to the width
of the distribution of the $D^{(q)}$ as displayed in Fig.~\ref{Dq_dist}. In the
next section we will show that this width is related to the quark number
susceptibility $\chi_q$. This is an extensive quantity, i.e., it grows with the
spatial volume $V = (aN_s)^3$. We thus conclude that the number of $D^{(q)}$ that
contribute significantly to the fugacity sum is proportional to the spatial volume $V$. 
In turn
this means that also $q_{cut}$ needs to be increased linearly in $V$.

\section{Quark number density and quark number susceptibility}

Let us finally discuss an exploratory
calculation of the quark number density and the quark number susceptibility based on canonical determinants and the fugacity expansion.
Combining the general expression for the grand canonical partition sum $Z(\mu)$ from
Eq.\ (\ref{zgrand}) with the fugacity expansion (\ref{detfugacity}) we find
\begin{eqnarray}
Z(\mu) & = & \int\!\! {\cal D}[U] \, e^{-S_G[U]} \! 
\left(\sum_q e^{q \mu\beta} \, D^{(q)} \right)^2 
\\
& = & \int\!\! {\cal D}[U] \, e^{-S_G[U]} \, \det[D(0)]^2 
\left(\!\sum_q  \, 
\frac{e^{q \mu\beta} \, D^{(q)}}{\det[D(0)]} \right)^{\!\!2} \!,
\nonumber
\end{eqnarray}
where in the second step we inserted $1 = \det[D(0)]^2/\det[D(0)]^2$ to write
the whole expression as a vacuum expectation value (up to normalization) in the
grand canonical ensemble at $\mu = 0$ where we perform our simulation. The definitions of the
quark number density $n_q$ and the quark number susceptibility $\chi_q$
are
\begin{equation}
\frac{n_q}{T^3} \, = \, \frac{\beta^3}{V} \frac{\partial \ln Z(\mu)}{\partial \mu \beta} 
 \; \; , \; \; \; \frac{\chi_q}{T^2} \, = \, \frac{\beta^3}{V} \frac{\partial^2 \ln Z(\mu)}{\partial (\mu \beta)^2} 
 \; .
\end{equation}
Both observables are intensive quantities after the normalization with the 3-volume $V$ and are made dimensionless using suitable powers of $T$. For the necessary first and second derivatives one finds  
\begin{eqnarray}
\frac{\partial \ln Z(\mu)}{\partial \mu \beta}  &\!\!=\!\!& \frac{2}{Z(\mu)}\! \int\!\! {\cal D}[U] \,
e^{-S_G[U]} 
\det[D(0)]^2 \; M_0  M_1 \; ,
\\
\frac{\partial^2\!\ln\!Z(\mu)}{\partial (\mu \beta)^2} &\!\!=\!\!& \frac{2}{Z(\mu)} \!\int\!\! {\cal D}[U] \, e^{-S_G[U]}  
\det[D(0)]^2\!\Big[\!M_2 M_0\!+\!M_1^2\! \Big] \nonumber
\\
 & & \hspace{5mm}
 - \, \left( \frac{\partial \ln Z(\mu)}{\partial \mu \beta} \right)^2 ,
\nonumber
\end{eqnarray}
where we introduced the moments $M_n, n = 0,1,2$ of the fugacity series as
\begin{equation}
M_n \; = \; \sum_{q= -q_{cut}}^{q_{cut}} \, q^n \, e^{q\mu\beta} \, \frac{D^{(q)}}{\det[D(0)]} \; .
\label{moments}
\end{equation}
Finally we express the normalization factor $1/Z(\mu)$ with the moment
$M_0$ and obtain (use  $\beta^3 / V = (N_t/N_s)^3$)
\begin{eqnarray}
n_q &\! = \! & 2\! \left(\frac{N_t}{N_s}\right)^{\!3} \frac{\langle \, M_0  M_1 \, \rangle}
{\langle \, (M_0)^2 \,\rangle} \; ,
\label{obsfinal}
\\ 
\chi_q &\! = \!& 2\! \left(\frac{N_t}{N_s}\right)^{\!3} \left[ \frac{\langle \, M_0  M_2 + (M_1)^2 \, \rangle}
{\langle \, (M_0)^2 \, \rangle} \, - \,  2 \left(\frac{\langle \, M_0  M_1 \, \rangle}
{\langle \, (M_0)^2 \,\rangle} \right)^{\!2\,} \right]\! .
\nonumber
\end{eqnarray}
In the final result (\ref{obsfinal}) both observables are expressed in terms of
vacuum expectation values of moments of the fugacity series. These vacuum
expectation values $\langle .. \rangle$ are computed with two flavors of
Wilson fermions at $\mu = 0$.  

\begin{figure}[t]
\begin{center}
\hspace*{-4mm}
\includegraphics[width=76mm,clip]{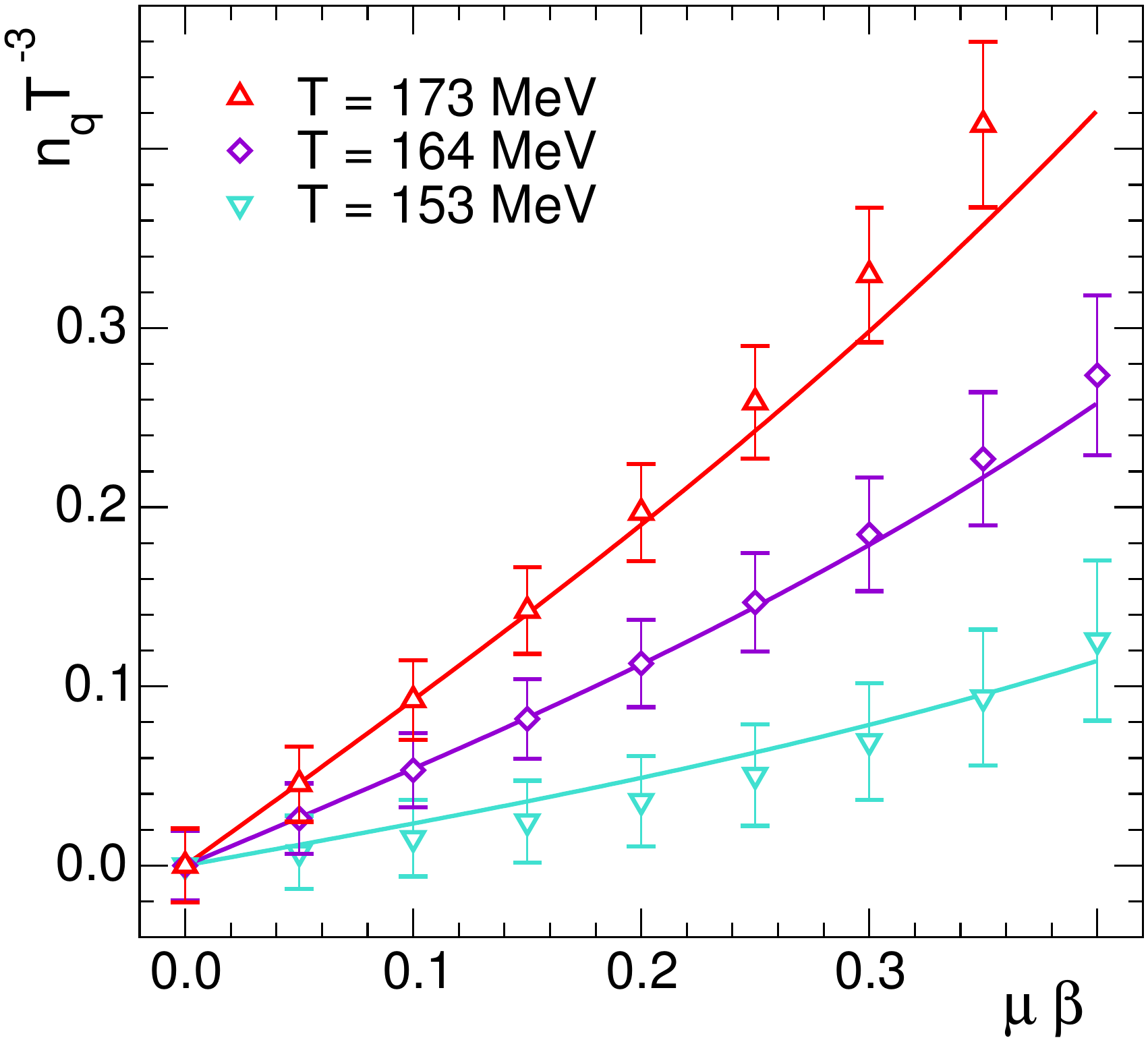} 
\end{center}
\vspace{-3mm}
\caption{Quark number density as a function of $\mu \beta$. We compare our results at three different temperatures
(symbols) to the results from Taylor expansion (curves).  \hfill} 
\label{nq} 
\end{figure}

In Fig.~\ref{nq} we show  our results (symbols) for $n_q/T^3$ as function of the dimensionless 
combination $\mu \beta$ for three different temperatures. We compare the results to the outcome of the Taylor expansion presented in \cite{taylor1,taylor2}. More precisely we used the terms up to fourth order from
\cite{taylor2} and interpolated the Taylor coefficients to the temperatures used in our study. However, this comparison should be viewed with caution: The calculation in \cite{taylor2} is done on considerably larger lattices, has a statistics that is by a factor 20 larger than the statistics available in our exploratory study, and also is based on the staggered formulation. Nevertheless we find reasonable agreement for our results at these temperatures and range of $\mu \beta$. 

\begin{table}[b]
\begin{center}
\begin{tabular}{ccc}
\quad  $T$  \quad & \quad Fugacity expansion \quad & \quad Taylor expansion \protect{\cite{taylor2}} \quad \\
\hline
144 MeV &  0.25(8) &  0.1360(44) \\
153 MeV &  0.14(9) &  0.2312(52) \\
157 MeV &  0.41(8) &  0.3190(58) \\
164 MeV &  0.52(7) &  0.5340(62) \\
173 MeV &  0.90(8) &  0.9170(66) \\
189 MeV &  3.41(8) &  1.3514(36) \\
211 MeV &  5.85(4) &  1.5240(30)
\end{tabular}
\end{center}
\vspace{-3mm}
\caption{Comparison of results for $\chi_q/T^2$ at $\mu = 0$.\hfill}
\label{chitable}
\end{table}

We also attempt a comparison of our results for the quark number susceptibility $\chi_q$ at $\mu = 0$ 
to the results from the Taylor expansion. The corresponding numbers are listed in Table~\ref{chitable} (again interpolated to our values of $T$).
We find that up to temperatures near the crossover the agreement of the two approaches is reasonable. 
Given the fact that the calculation \cite{taylor2} and our exploratory study differ considerable in statistics and volume, this agreement is satisfactory. Above the crossover we see, however, a rather strong discrepancy. We suspect two main reasons for this discrepancy: 1) The rather small volumes that are available for our calculation, and 2) the fact that for the larger values of $T$ higher Fourier modes contribute significantly (the factor $q^2$ in the second moment $M_2$ of Eq.~(\ref{moments}) enhances them further) which in the current setting of 256 sampling points for the numerical evaluation of the Fourier integral are not sufficiently accurate.
\relax

\section{Summary and discussion}

In this exploratory study we analyzed the canonical determinants of 2-flavor lattice QCD with 
Wilson fermions. The canonical determinants are the coefficients in the fugacity expansion of the 
fermion determinant and may be calculated as the Fourier modes with respect to imaginary 
chemical potential. We speed up this evaluation by using a dimensional reduction formula for the grand canonical determinant. 

We illustrate that a sizable dependence of the fermion determinant on the imaginary chemical potential sets in only 
at temperatures near the crossover temperature, which already shows that at the crossover an enhancement 
of the higher quark numbers can be expected. Studying the size distribution of the canonical determinants we find 
that their distribution is roughly Gaussian in the net quark number $q$ with a width that starts to increase at the 
crossover. The average  phases of the canonical determinants drop quickly with $q$, but the drop is slowed above 
the crossover temperature. This analysis sheds light on the complex phase problem from the point of view of the 
fugacity series. 

We continue with a systematical analysis of the convergence properties of the fugacity expansion by comparing 
the truncated fugacity series to an exact evaluation of the grand canonical determinant at several values of the 
chemical potential. It is shown that for lower temperatures we can obtain very good accuracy up to $\mu \beta \sim 
0.8$, while above the crossover we only reach $\mu \beta \sim 0.4$. We stress that these results are specific for 
the numerical effort we invest here, in particular 256 sampling points for the evaluation of the Fourier moments. 
On a  finite lattice the fugacity expansion is a finite Laurent series, and in principle it is possible to compute all 
coefficients such that the series representation of the grand canonical determinant becomes exact. 

Finally we use the fugacity series to explore the evaluation of the quark number
density and the quark number  susceptibility through first and second moments of
the truncated fugacity series. This is an interesting alternative to  a standard
calculation where the quark number $q$ is a binomial in the quark fields, $q
\propto \int d^4 x \overline{\psi}(x) \gamma_4  \psi(x)$ which after integrating
out the fermions is related to the trace of the quark propagator, $q \propto$ Tr
$ \gamma_4 D^{-1}$. In this form $q$ is difficult to evaluate numerically using,
e.g., stochastic estimators. Once the canonical determinants $D^{(q)}$ are
available  the expressions in terms of the moments of the fugacity expansion are
considerably simpler and we demonstrate  that the approach is compatible with
the conventional method up to temperatures where we have sufficiently  control
over the fugacity series. 

Conceptually the fugacity expansion falls in the same category as the
Taylor expansion: A series is used to extrapolate the information from a Monte
Carlo simulation at $\mu = 0$ to finite values of the chemical potential. Thus
both, Taylor- and fugacity expansion face the same overlap problem, i.e., the
$\mu = 0$ configurations used for evaluating the expansion coefficients may have
rather little overlap with the configurations that dominate physics at finite
$\mu$. This limits both expansion approaches to small values of $\mu$. 

Concerning other 
properties, Taylor- and fugacity expansion are different: At finite volume the
fugacity series is a finite Laurent series in the fugacity parameter $e^{\mu \beta}$, while the Taylor series is an infinite series even at finite volume. Thus
the two series will have different properties and one should explore which of
the two approaches provides a better expansion around $\mu = 0$. To study this
question we currently compare the two series in an effective theory for QCD,
where a flux representation free of the complex phase problem 
allows one to obtain high precision reference data from a Monte Carlo 
simulation \cite{flux}. These will be used to study the quality of the
approximation from Taylor- and fugacity series. 

We conclude with stressing that this exploratory work should be considered only
as a first step towards a full development of fugacity expansion as a reliable
tool in finite density lattice QCD. Certainly further improvement of the
numerical techniques and more tests will be necessary to assess the full
potential of the approach.

\vspace{7mm}
\noindent									
{\bf Acknowledgments:} We thank Christian Lang, Anyi Li and Bernd-Jochen Schaefer for
interesting discussions and remarks. We also would like to thank Christian Lang and
Ludovit Liptak for help with parts of the computer code. This work has been supported by
the Austrian Science Fund FWF: Doctoral School ''Hadrons in Vacuum, Nuclei and Stars'', DK W1203. 
Christof Gattringer thanks the members of the Institute for Nuclear Theory at the  University of Washington, 
Seattle, where part of this work was done, for their hospitality and the inspiring atmosphere.
Christof Gattringer also acknowledges support by the Dr.\ Heinrich-J\"org foundation of 
the University of Graz. The simulations were done
on the Linux clusters of the computer center at the University of Graz, and we thank
their staff for service and support. This work was in part based on the MILC collaboration's
public lattice gauge theory code.

\clearpage
\end{document}